# On Testing Unitarity of the Quark Mixing Matrix [1]

Zhi-zhong XING [2]

*Sektion Physik, Theoretische Physik, Universität München,*
*Theresienstrasse 37, D-80333 München, Germany*

## Abstract

Unitarity triangles and characteristic measurables of the $3 \times 3$ Cabibbo-Kobayashi-Maskawa (CKM) matrix are discussed. Beyond the $3 \times 3$ CKM scheme we make a rephasing-invariant generalization of the Gronau-Wyler-Dunietz approach to determine the weak phase shift in $B_u^\pm \rightarrow DK^\pm$ or $B_d^0 \rightarrow DK^{*0}$ vs $\bar{B}_d^0 \rightarrow D\bar{K}^{*0}$, which is only sensitive to the underlying new physics in $D^0 - \bar{D}^0$ mixing. We also show that the weak angle $\gamma \equiv \arg(-V_{ub}^* V_{cd}^* V_{ud} V_{cb})$ is possible to be determined from the $CP$ asymmetries of some $B_d$ decays, even in a non-standard model with an extended quark sector. Brief comments are given on tests of unitarity of the $3 \times 3$ CKM matrix.

---

# Part I. Within the $3 \times 3$ CKM Scheme

In this part we explore some consequences of unitarity of the $3 \times 3$ CKM matrix. An instructive discussion is given about the characteristic measurables of the CKM matrix and their relations with the unitarity triangles. We also take a look into the feature of the weak angle $\gamma$ in exclusive $B$ decays and $CP$ asymmetries.

## A. Unitarity Triangles and Parametrizations

In the minimal standard electroweak model, the $3 \times 3$ CKM matrix $V$ provides a natural description of quark mixing and $CP$ violation. Unitarity is the only but powerful constraint, imposed by the model itself, on $V$. This restriction can be expressed as two sets of orthogonality-plus-normalization conditions:

$$\sum_{\alpha=d,s,b} (V^*_{i\alpha} V_{j\alpha}) = \delta_{ij} , \qquad \sum_{i=u,c,t} (V^*_{i\alpha} V_{i\beta}) = \delta_{\alpha\beta} , \qquad (1)$$

where Latin subscripts run over the up-type quarks $(u, c, t)$ and Greek ones over the down-type quarks $(d, s, b)$. In the complex plane the six orthogonality relations given above correspond to six triangles (see Fig. 1), the so-called unitarity triangles.

By use of the unitarity conditions in Eq. (1), one is able to parametrize the CKM matrix in various ways. Several popular parametrizations are given in terms of three Euler angles and one $CP$-violating phase [1, 2]. A parametrization is also available in terms of four independent modulus of the CKM matrix elements [3], or four independent sides of the unitarity triangles, or four independent angles of the unitarity triangles [4]. To analyze data, the Wolfenstein parametrization is most convenient because it straightforwardly reflects the hierarchy of quark mixings [5]. However, the unitarity conditions in the original Wolfenstein form are satisfied only to the accuracy of $O(\lambda^4)$, which is insufficient for a self-consistent description of all properties of the CKM matrix. A useful modified version can be given as [6]

$$V = \begin{pmatrix} 1 - \frac{1}{2}\lambda^2 - \frac{1}{8}\lambda^4 & \lambda & A\lambda^3(\rho - i\eta) \\ -\lambda \left[1 + \frac{1}{2}A^2\lambda^4(2\rho - 1) + iA^2\lambda^4\eta\right] & 1 - \frac{1}{2}\lambda^2 - \frac{1}{8}(4A^2 + 1)\lambda^4 & A\lambda^2 \\ A\lambda^3(1 - \rho - i\eta) & -A\lambda^2 \left[1 + \frac{1}{2}\lambda^2(2\rho - 1) + i\lambda^2\eta\right] & 1 - \frac{1}{2}A^2\lambda^4 \end{pmatrix} , \quad (2)$$

where unitarity is kept up to $O(\lambda^6)$. This degree of accuracy is enough for $V$ to confront all precise experimental data in the near future.

All measurables of $CP$ violation in the standard model are sensitively related to the angles of the unitarity triangles. It is instructive to express the nine inner angles in terms of the Wolfenstein

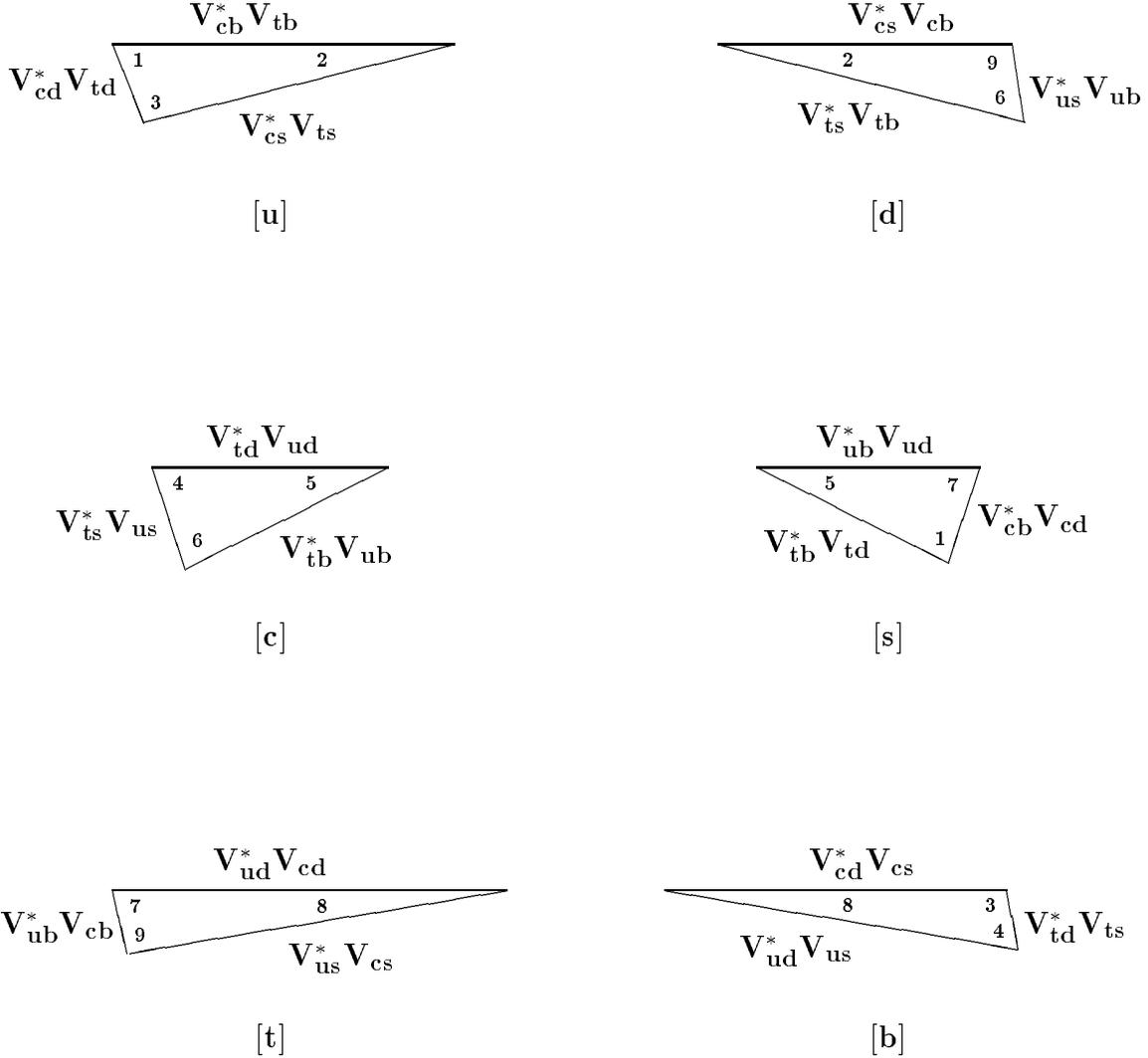

Figure 1: The unitarity triangles of the 3 × 3 CKM matrix in the complex plane. Each triangle is named in terms of the quark flavor that does not appear in its three sides. Note that the six triangles have the same area, and they only have nine different inner angles (versus eighteen different sides).

parameters. In lowest-order approximations, we obtain:

$$\tan(\angle 1) \approx \tan(\angle 4) \approx -\tan(\angle 3) \approx \frac{\eta}{1-\rho}, \tag{3a}$$

$$\tan(\angle 6) \approx \tan(\angle 7) \approx -\tan(\angle 9) \approx \frac{\eta}{\rho}, \tag{3b}$$

and

$$\tan(\angle 2) \approx \lambda^2 \eta, \qquad \tan(\angle 8) \approx A^2 \lambda^4 \eta, \qquad \tan(\angle 5) \approx \frac{\eta}{\rho(\rho-1)+\eta^2}. \tag{3c}$$

It is clear that all the six triangles collapse into lines if $\eta = 0$. Conventionally one uses $\alpha = \angle 5$, $\beta = \angle 1$, and $\gamma = \angle 7$ to denote the three angles of unitarity triangle [s], which will be overdetermined at $B$-meson factories. Here it is worth emphasizing that the angles of the other unitarity triangles also have chances to be determined with the development of more precise experiments. We expect that the approximate relations given in Eq. (3) can be tested by various measurements of $CP$ violation in the near future, either within or beyond the $K$-, $D$- and $B$-meson systems.

## B. Characteristic Measurables

The 3×3 CKM matrix has four independent characteristic measurables. The first one is a universal measure of $CP$ violation in weak interactions of quarks, the so-called Jarlskog parameter $\mathcal{J}$ [7]:

$$\mathcal{J} = \left| \mathrm{Im}\left( V_{i\alpha} V_{j\beta} V_{i\beta}^* V_{j\alpha}^* \right) \right| \qquad (i \neq j, \ \alpha \neq \beta). \tag{4}$$

It is straightforward to show that all the six unitarity triangles have the same area $\mathcal{J}/2$, although their shapes are quite different (see Fig. 1 for illustration). If there is no $CP$ violation, i.e., $\mathcal{J} = 0$, then all six unitarity triangles collapse into lines. An interesting point is that $\mathcal{J}$ can be determined from three sides of each triangle, which are not directly related to any $CP$-violating signal.

The structure of the 3×3 CKM matrix is basically characterized by its two off-diagonal asymmetries [8]. They are denoted by $\mathcal{A}_1$ about the $V_{ud} - V_{cs} - V_{tb}$ axis [9] and $\mathcal{A}_2$ about the $V_{ub} - V_{cs} - V_{td}$ axis (see Fig. 2 for illustration):

$$\mathcal{A}_1 = |V_{us}|^2 - |V_{cd}|^2 = |V_{cb}|^2 - |V_{ts}|^2 = |V_{td}|^2 - |V_{ub}|^2, \tag{5a}$$

$$\mathcal{A}_2 = |V_{us}|^2 - |V_{cb}|^2 = |V_{cd}|^2 - |V_{ts}|^2 = |V_{tb}|^2 - |V_{ud}|^2. \tag{5b}$$

The above relations are direct consequences of the normalization conditions given in Eq. (1). Note that the asymmetry parameters $\mathcal{A}_{1,2}$ are independent of each other, and they are independent of the $CP$-violating parameter $\mathcal{J}$.

Note that the element $V_{cs}$ sits at the centre of the CKM matrix and is independent of the off-diagonal asymmetries $\mathcal{A}_{1,2}$ (see Fig. 2). $|V_{cs}|$ is indeed the fourth characteristic measurable of the 3×3

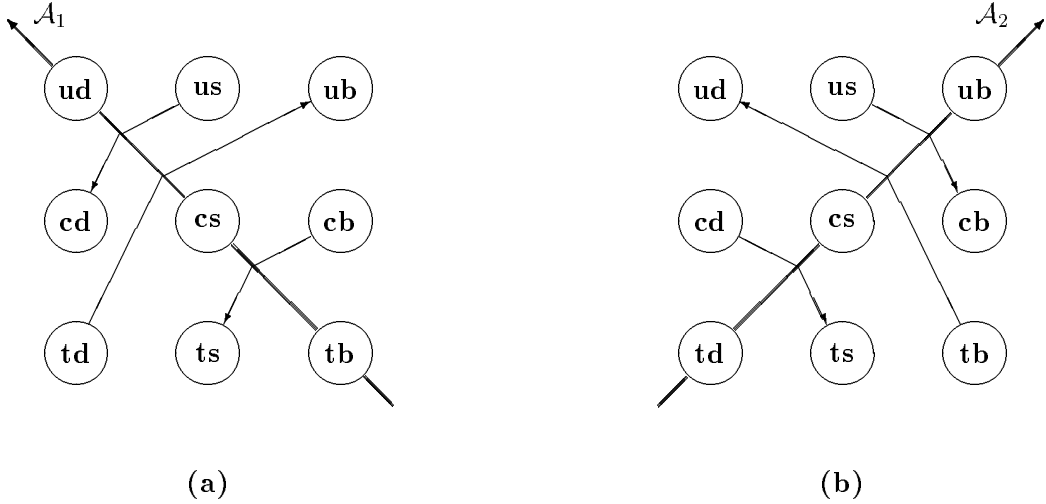

Figure 2: The off-diagonal asymmetries of the $3\times 3$ CKM matrix: (a) $\mathcal{A}_1$ about the $V_{ud}-V_{cs}-V_{tb}$ axis; and (b) $\mathcal{A}_2$ about the $V_{ub}-V_{cs}-V_{td}$ axis.

CKM matrix [8]. In contrast, either $2\times 2$ or $4\times 4$ unitary matrix has not such a "central" element to characterize its geometrical structure. Of course, one can use $\mathcal{J}, \mathcal{A}_{1,2}$ and $|V_{cs}|^2$ to parametrize the whole CKM matrix $V$. In such a parametrization, the matrix elements $|V_{us}|, |V_{cd}|, |V_{cb}|$ and $|V_{ts}|$ are independent of the $CP$-violating parameter $\mathcal{J}$, while $|V_{ud}|, |V_{ub}|, |V_{td}|$ and $|V_{tb}|$ depend upon all the four parameters.

By use of the modified Wolfenstein parametrization in Eq. (2), $\mathcal{J}$ and $\mathcal{A}_{1,2}$ are given as [8]

$$\mathcal{J} \approx A^2\lambda^6\eta, \qquad \mathcal{A}_1 \approx A^2\lambda^6(1-2\rho), \qquad \mathcal{A}_2 \approx \lambda^2\left(1 - A^2\lambda^2\right). \qquad (6)$$

It is clear that $\mathcal{A}_2 >> \mathcal{A}_1$ and $\mathcal{A}_1 \sim \mathcal{J}$. Both $\mathcal{A}_1$ and $\mathcal{A}_2$ are independent of $\eta$, a parameter signifying $CP$ violation. The allowed ranges of the Wolfenstein parameters have been analyzed by many authors with the help of current experimental data. Taking only the central values of $\lambda, A, \rho$ and $\eta$ into account [10], we find $\mathcal{A}_2/\mathcal{A}_1 \geq 400$, $\mathcal{A}_1 \sim 10^{-5} - 10^{-4}$, and $\mathcal{J} \sim 10^{-5}$.

From the direct measurements $|V_{us}| = 0.2205 \pm 0.0018$ and $|V_{cd}| = 0.204 \pm 0.017$ [1], we observe that the possibility of $\mathcal{A}_1 \approx 0$ has not been completely ruled out. But one is convinced that $\mathcal{A}_1 > 0$ should be of the dominant possibility. If this point is really true, then we can find a definite hierarchy for the nine CKM matrix elements:

$$\begin{aligned} |V_{tb}| > |V_{ud}| > |V_{cs}| &>> |V_{us}| > |V_{cd}| \\ &>> |V_{cb}| > |V_{ts}| \\ &>> |V_{td}| > |V_{ub}|. \end{aligned} \qquad (7)$$

Since $|V_{ub}/V_{cb}| = 0.08 \pm 0.02$ [1], it is certain that all $|V_{i\alpha}|$ are nonzero. The above interesting result reflects our present understanding of the magnitudes of quark mixings.

Geometrically the nonvanishing $\mathcal{A}_1$ and $\mathcal{A}_2$ imply some differences in the six unitarity triangles. In general, these triangles have nine different inner angles and eighteen different sides, thus their shapes are different from one another (see Fig. 1). If $\mathcal{A}_1$ or $\mathcal{A}_2$ were vanishing, one would find three equivalence relations among the six triangles [8]:

$$\mathcal{A}_1 \;=\; 0 \quad \Longrightarrow \quad [\mathbf{u}] \cong [\mathbf{d}]\,, \qquad [\mathbf{c}] \cong [\mathbf{s}]\,, \qquad [\mathbf{t}] \cong [\mathbf{b}] \tag{8a}$$

with $\angle 1 = \angle 6$, $\angle 3 = \angle 9$, $\angle 4 = \angle 7$; and

$$\mathcal{A}_2 \;=\; 0 \quad \Longrightarrow \quad [\mathbf{u}] \cong [\mathbf{b}]\,, \qquad [\mathbf{c}] \cong [\mathbf{s}]\,, \qquad [\mathbf{t}] \cong [\mathbf{d}] \tag{8b}$$

with $\angle 1 = \angle 4$, $\angle 2 = \angle 8$, $\angle 6 = \angle 7$. In either case, the six unitarity triangles have six different inner angles and nine different sides. As a consequence, the CKM matrix can be parametrized by use of three independent quantities. In view of Eq. (6), there is no possibility for $\mathcal{A}_2 = 0$. The possibility of $\mathcal{A}_1 = 0$, which requires $\rho \approx 0.5$, is only allowed on the extreme margin of the existing data and should be absolutely ruled out in the near future. From the point of view that the quark mixing matrix $V$ is basically governed by the quark mass matrices $M_\mathrm{U}$ and $M_\mathrm{D}$ [11], we stress that the nonzero off-diagonal asymmetries of $V$ would imply a kind of symmetry breaking appearing in $M_\mathrm{U,D}$. Thus it is suggestive that the study of specific patterns of $M_\mathrm{U,D}$ (and their underlying dynamics) may start from the symmetry limit $\mathcal{A}_1 = 0$ (or $\mathcal{J} = 0$) at a superheavy scale.

## C. Triangle [s] and Angle $\gamma$

Among six unitarity triangles of the $3 \times 3$ CKM matrix, triangle [$\mathbf{s}$] is of most interest for studies of $CP$ violation in $B$-meson systems. Its three inner angles are commonly denoted by Greek letters $\alpha$, $\beta$ and $\gamma$ (see Fig. 3 for illustration):

$$\alpha \;=\; \arg\left(-\frac{V_{tb}^*V_{td}}{V_{ub}^*V_{ud}}\right)\,, \qquad \beta \;=\; \arg\left(-\frac{V_{cb}^*V_{cd}}{V_{tb}^*V_{td}}\right)\,, \qquad \gamma \;=\; \arg\left(-\frac{V_{ub}^*V_{ud}}{V_{cb}^*V_{cd}}\right)\,. \tag{9}$$

Of course,

$$\alpha \;+\; \beta \;+\; \gamma \;=\; 180^0 \tag{10}$$

is a trivial consequence of the above definition, no matter whether the vectors $V_{ub}^*V_{ud}$, $V_{cb}^*V_{cd}$ and $V_{tb}^*V_{td}$ form a closed triangle or not in the complex plane. This point can be seen more clearly in the following sections. Hence an experimental examination of the sum rule in Eq. (10) does not make much sense for testing unitarity of the $3 \times 3$ CKM matrix. In the literature there are still some ambiguities or misleading remarks associated with this problem. It is therefore necessary to make a clarification.

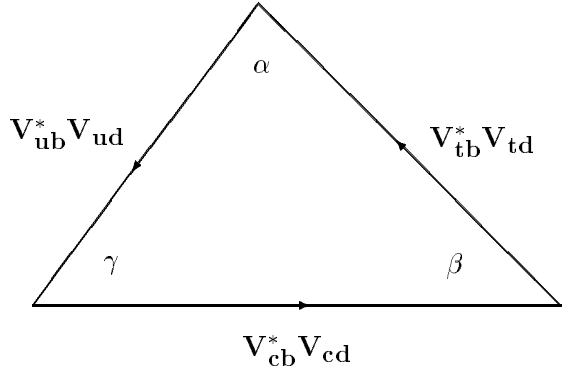

Figure 3: Unitarity triangle [s].

It is known that the angles $\alpha$ and $\beta$ can directly enter the $CP$ asymmetries of some $B_d$ transitions, which are induced by the interplay of decay and $B_d^0 - \bar{B}_d^0$ mixing (see, e.g., Ref. [12]). The typical examples are $B_d^0$ vs $\bar{B}_d^0 \to \psi K_S$ and $\pi^+\pi^-$, whose $CP$ asymmetries are governed by $\beta$ and $\alpha$ respectively. Taking into account the penguin contributions to $B_d \to \pi^+\pi^-$, one may extract $\alpha$ from the correlative decay modes $B_d \to \pi^+\pi^-, \pi^0\pi^0$ and $B_d^\pm \to \pi^\pm\pi^0$ with the help of isospin relations and time-dependent measurements [13].

Since the CKM factors $V_{ub}^*V_{ud}$ and $V_{cb}^*V_{cd}$ come from the quark processes $\bar{b} \to \bar{u}u\bar{d}$ and $\bar{b} \to \bar{c}c\bar{d}$ respectively, their relative phase $\gamma$ cannot be related to any $CP$-violating observable of $B_u$, $B_d$ or $B_s$ decays *in an exact way*. All the proposed approaches for extraction of $\gamma$ are only able to probe the approximate magnitude of $\gamma$, and they may work to a good degree of accuracy only in the assumption of unitarity of the $3 \times 3$ CKM matrix. For example, the $CP$ asymmetry of $B_s \to \rho^0 K_S$ is indeed governed by the phase of $(V_{tb}^*V_{ud}^*V_{cs}^*V_{ts}V_{ub}V_{cd})$ other than that of $(-V_{ub}^*V_{cd}^*V_{ud}V_{cb})$, but the two phases are approximately identical if triangle [d] closes and its side $|V_{us}^*V_{ub}|$ is vanishingly short [14]. In and only in weak $B$ decays of the type

$$\begin{aligned} B_c^+ &\longrightarrow (\pi^+, \rho^+, a_1^+) + (D^0, D^{*0}), \\ B_c^+ &\longrightarrow (\pi^0, \rho^0, a_1^0, \omega) + (D^+, D^{*+}), \end{aligned} \quad (11)$$

a pure $\gamma$ can enter the decay rates. These decays occur through the spectator diagrams ($\bar{b} \to \bar{u}u\bar{d}$), the annihilation diagrams ($\bar{b} \to \bar{c}c\bar{d}$) and the penguin diagrams ($\bar{b} \to \bar{d}$). Due to the orthogonality condition performed in triangle [s], one of the three CKM factors in the penguin amplitudes (i.e., $V_{tb}^*V_{td}$) can be absorbed. Thus the overall amplitude of every decay mode in Eq. (11) contains two components, associated with $V_{ub}^*V_{ud}$ and $V_{cb}^*V_{cd}$ respectively. For instance, the $CP$ asymmetry in $B_c^+ \to \pi^+ D^0$ vs $B_c^- \to \pi^- \bar{D}^0$ is unambiguously proportional to $\sin\gamma$. But this asymmetry involves large uncertainty from final-state strong interactions [15], hence in practice the above $B_c^\pm$ transitions cannot be used to extract $\gamma$.

Gronau and Wyler have developed an approach to extract $\gamma$ from the decay modes $B_u^\pm \to D_{1(2)}^0 K^\pm$, where $D_{1(2)}^0 = [D^0 + (-)\bar{D}^0]/\sqrt{2}$ denotes a $CP$ even (odd) state [16]. Following the same idea, Dunietz carried out an analysis of $B_d^0 \to D_{1(2)}^0 K^{*0}$ vs $\bar{B}_d^0 \to D_{1(2)}^0 \bar{K}^{*0}$ [17]. Note that the relevant weak phase shift in these transitions is indeed

$$\gamma_0 \;=\; \arg\left(\frac{V_{ub}^* V_{cs}}{V_{cb}^* V_{us}} \cdot \frac{V_{us} V_{cs}^*}{V_{us}^* V_{cs}}\right) \;=\; \arg\left(\frac{V_{ub}^* V_{us}}{V_{cb}^* V_{cs}}\right) \tag{12}$$

other than $\gamma$. In the above formula we have taken into account the weak phases from $D^0 - \bar{D}^0$ mixing, otherwise, $\gamma_0$ is *not* rephasing-invariant. However, $\gamma_0 \approx \gamma$ is a very good approximation within the $3 \times 3$ CKM scheme. This point can be clearly seen as follows. From the orthogonality relation shown by triangle [t], we find that $V_{ud}^* V_{cd} \approx -V_{us}^* V_{cs}$ holds up to the accuracy of $O(\lambda^5)$. Hence one obtains $\gamma \approx \arg(V_{ub}^* V_{cs}^* V_{cb} V_{us})$, equal to $\gamma_0$ in Eq. (12). It should be noted that $\gamma_0 \approx \gamma$ will not be valid if unitarity of the $3 \times 3$ CKM matrix is violated. In this case, an apparent weak phase from new physics may enter $\gamma_0$ through $D^0 - \bar{D}^0$ mixing.

# Part II.   Beyond the $3 \times 3$ CKM Scheme

In this part we first make a rephasing-invariant generalization of the Gronau-Wyler-Dunietz approach to determine a weak phase beyond the $3 \times 3$ CKM scheme. In either the model of four quark families or that of $Z$-mediated flavor changing neutral currents (FCNC's), we show that $\gamma$ is possible to be determined from $CP$ asymmetries of some $B_d$ decays with the help of Eq. (10), although both $\alpha$ and $\beta$ may be significantly contaminated by new physics. Finally we comment briefly on tests of unitarity of the $3 \times 3$ CKM matrix.

## D.   Determination of a Weak Phase

Before the $3 \times 3$ CKM mechanism passes stringent tests, it is useful to develop some model-independent approaches for extraction of the weak phases from specific $B$-meson decays. Such ideas rely on the fact that there is no significant effect of new physics on the direct decay of $b$ quark via the tree-level $W$-mediated diagrams [12]. Violation of the unitarity conditions in Eq. (1) mainly manifests itself in $B^0 - \bar{B}^0$ mixing (or $D^0 - \bar{D}^0$ mixing) and loop-induced penguin channels. To illustrate, we make a generalization of the Gronau-Wyler-Dunietz approach to determine a weak phase from the decay modes $B_u^\pm \to D K^\pm$ or from $B_d^0 \to D K^{*0}$ vs $\bar{B}_d^0 \to D \bar{K}^{*0}$. This weak phase can be denoted, generally and rephasing-invariantly, as

$$\varphi \;=\; \arg\left(\frac{V_{ub}^* V_{cs}}{V_{cb}^* V_{us}} \cdot \frac{q_D}{p_D}\right) \;, \tag{13}$$

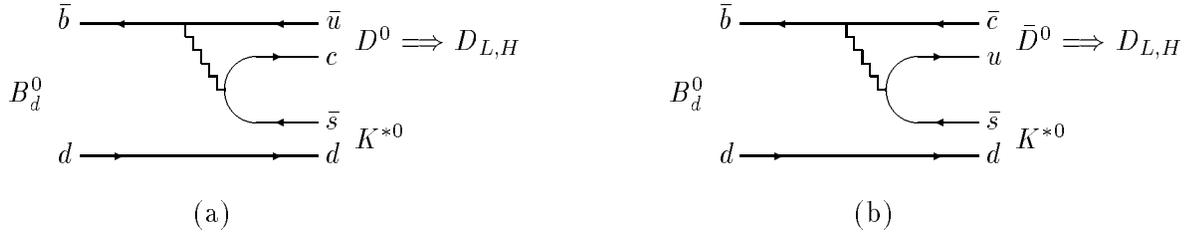

Figure 4: The dominant tree-level $W$-mediated diagrams for $B_d^0 \to D_L K^{*0}$ or $D_H K^{*0}$.

where the complex parameters $q_D$ and $p_D$ connect $D^0$ and $\bar{D}^0$ to their mass eigenstates through

$$\begin{aligned} |D_L\rangle &= p_D|D^0\rangle + q_D|\bar{D}^0\rangle \\ |D_H\rangle &= p_D|D^0\rangle - q_D|\bar{D}^0\rangle \end{aligned} \quad (14)$$

(with $|p_D|^2 + |q_D|^2 = 1$). As pointed out by Blaylock, Seiden and Nir [18], $|q_D/p_D| \approx 1$ is a very reasonable approximation in all reasonable models of $D^0 - \bar{D}^0$ mixing with new physics. This implies that $CP$ violation in the $D^0 - \bar{D}^0$ mass matrix is negligible. In the minimal standard model, $q_D/p_D = (V_{us}V_{cs}^*)/(V_{us}^*V_{cs})$, $\varphi$ turns out to be $\gamma_0$ as given by Eq. (12). For some non-standard models like those listed in Ref. [18], however, $q_D/p_D$ may introduce a significant phase shift into $\varphi$.

Let us take the decay modes $B_d^0 \to DK^{*0}$ and $\bar{B}_d^0 \to D\bar{K}^{*0}$ for example. Since $B_d^0 \to D^0 K^0$, $B_d^0 \to \bar{D}^0 K^0$ and their $CP$-conjugate counterparts occur only through the tree-level $W$-mediated quark diagrams (see Fig. 4 for illustration), we parametrize the transition amplitudes of $B_d \to D_{L(H)} K^*$ as follows:

$$\begin{aligned} A(B_d^0 \to D_{L(H)}K^{*0}) &= p_D^* \, (V_{ub}^* V_{cs}) \, A_a \, e^{i\delta_a} \, +(-) \, q_D^* \, (V_{cb}^* V_{us}) \, A_b \, e^{i\delta_b} \,, \\ A(\bar{B}_d^0 \to D_{L(H)}\bar{K}^{*0}) &= p_D^* \, (V_{cb} V_{us}^*) \, A_b \, e^{i\delta_b} \, +(-) \, q_D^* \, (V_{ub} V_{cs}^*) \, A_a \, e^{i\delta_a} \,, \end{aligned} \quad (15)$$

where $A_a$ and $A_b$ are real (positive) hadronic matrix elements, and $\delta_a$ and $\delta_b$ are the corresponding strong phases. Unlike ref. [17], here one cannot use a simple triangular relation to describe the above decay amplitudes. Some specific measurements are possible to establish the following (dimensionless) decay-rate asymmetry:

$$\Delta_{ij} \equiv \frac{|A(B_d^0 \to D_i K^{*0})|^2 - |A(\bar{B}_d^0 \to D_i \bar{K}^{*0})|^2}{|A(B_d^0 \to D^0 K^{*0})| \, |A(B_d^0 \to \bar{D}^0 K^{*0})|} \quad (16)$$

with $i,j = L$ or $H$. Denoting the strong phase difference $\delta_b - \delta_a \equiv \delta$ and using the reasonable approximation $|q_D/p_D| \approx 1$, we explicitly obtain

$$\begin{aligned} \Delta_{LL} &= 2\sin\varphi \, \sin\delta \,, & \Delta_{HH} &= -2\sin\varphi \, \sin\delta \,, \\ \Delta_{LH} &= 2\cos\varphi \, \cos\delta \,, & \Delta_{HL} &= -2\cos\varphi \, \cos\delta \,. \end{aligned} \quad (17)$$

In experiments, the relations $\Delta_{LL} = -\Delta_{HH}$ and $\Delta_{LH} = -\Delta_{HL}$ can be well examined. Note that only the asymmetries $\Delta_{LL}$ and $\Delta_{HH}$ represent $CP$ violation, and they vanish if the weak phase shift $\varphi$ vanishes.

Obviously Eq. (17) can be used to extract $\varphi$. If the $CP$ asymmetries $\Delta_{LL}$ and $\Delta_{HH}$ were substantially suppressed due to the smallness of $\delta$, then $\Delta_{LH} = -\Delta_{HL} \approx 2\cos\varphi$ would be a good approximation. In general, we have

$$\left(\frac{\Delta_{LL}}{\sin\varphi}\right)^2 + \left(\frac{\Delta_{LH}}{\cos\varphi}\right)^2 = \left(\frac{\Delta_{LL}}{\sin\delta}\right)^2 + \left(\frac{\Delta_{LH}}{\cos\delta}\right)^2 = 4 \; . \tag{18}$$

Note that the angle $\varphi$ (or $\delta$) extracted from the above equation has a few ambiguities in its size and sign. This kind of ambiguities can be removed by studying a set of exclusive decay modes $B_d \to (D^0, \bar{D}^0, D_L, D_H) + X^0$, where $X^0$ is any mode with flavor content $(\bar{s}d)$ or $(s\bar{d})$, as long as its net strangeness can be unambiguously deduced [17]. All such processes have a common weak phase shift $\varphi$, but their strong phase shifts $\delta$ should be different from one another.

In a similar way, one can make a rephasing-invariant generalization of Gronau and Wyler's work in Ref. [16], so as to extract the weak phase shift $\varphi$ from the processes $B_u^\pm \to D_{L(H)} K^\pm$, etc [19].

## E. $\gamma$ in two Models with an Extended Quark Sector

A violation of unitarity of the $3 \times 3$ CKM matrix implies

$$\Omega \equiv V_{ub}^* V_{ud} + V_{cb}^* V_{cd} + V_{tb}^* V_{td} \neq 0 \; , \tag{19}$$

i.e., the three sides of triangle [s] do not close. In this case, $V_{ub}^* V_{ud}$, $V_{cb}^* V_{cd}$, $V_{tb}^* V_{td}$ and $-\Omega$ form a quadrangle in the complex plane (see Fig. 5 for illustration). Note that the angle sum rule in Eq. (10) remains valid, although the magnitudes of $\alpha$, $\beta$ and $\gamma$ all have been contaminated by new physics. Beyond the minimal standard model, here we consider two basic approaches to extend the quark sector [12, 20], which allow breaking of the unitarity constraints in Eq. (1):

(1) In the standard model with four quark families, quark mixing is described by a $4 \times 4$ unitary matrix. Thus $\Omega = -V_{t'b}^* V_{t'd}$, where $t'$ denotes the fourth up-type quark. The $t'$ quark can contribute to $B_d^0 - \bar{B}_d^0$ mixing via box diagrams, proportional to $(V_{t'b} V_{t'd}^*)^2$.

(2) In the model with an iso-singlet down-type quark, $\Omega = U_{db}$, where $U_{db}$ is a non-diagonal coupling of the $Z$ gauge boson. There may be tree-level $Z$-mediated FCNC's contributing to direct $b$ decays, but they are negligibly small in comparison with the tree-level $W$-mediated channels. However, $B_d^0 - \bar{B}_d^0$ mixing is possible to be significantly modified by tree-level $Z$-mediated diagrams, proportional to $(U_{db})^2$.

For both cases, the presence of $\Omega$ induces a new weak phase $\Phi$ to the $B_d^0 - \bar{B}_d^0$ mixing parameter of the minimal standard model:

$$\frac{q_B}{p_B} = \frac{V_{tb}^* V_{td}}{V_{tb} V_{td}^*} \quad \Longrightarrow \quad \frac{q_B}{p_B} = \frac{V_{tb}^* V_{td}}{V_{tb} V_{td}^*} e^{i\Phi} \; , \tag{20}$$

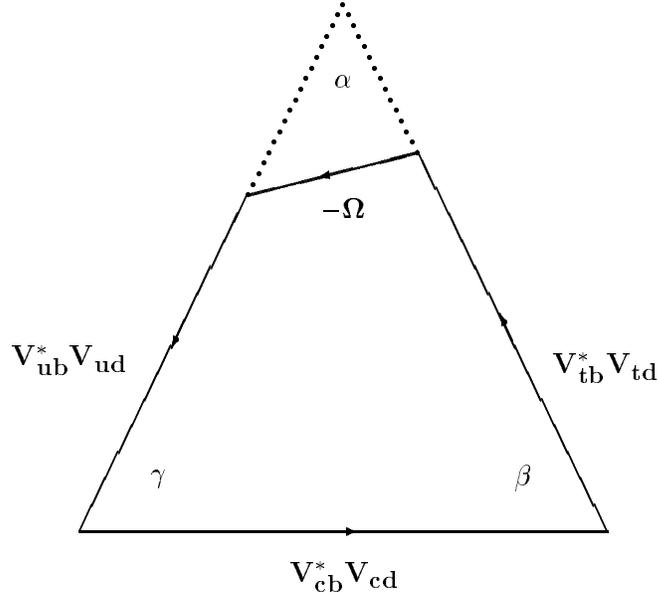

Figure 5: The unitarity quadrangle in the complex plane.

where
$$\Phi \;=\; \arg\left[1 \;+\; 2\,\frac{V_{t'b}^* V_{t'd}}{V_{tb}^* V_{td}}\,\frac{E(t,t')}{E(t,t)} \;+\; \frac{(V_{t'b}^* V_{t'd})^2}{(V_{tb}^* V_{td})^2}\,\frac{E(t',t')}{E(t,t)}\right] \tag{21a}$$

for the model of four quark families; and

$$\Phi \;=\; \arg\left[1 \;+\; \frac{U_{db}^2}{(V_{tb}^* V_{td})^2}\,\frac{4\pi \sin^2\theta_{\mathrm{w}}}{\alpha_e E(t,t)}\right] \tag{21b}$$

for the model of $Z$-mediated FCNC's. In the above equations, $E(i,j)$ denotes the box-diagram function with internal $i$ and $j$ quarks [21], $\alpha_e$ and $\theta_{\mathrm{w}}$ are the standard electroweak parameters.

Let us show how $\Phi$ enters the $CP$ asymmetries induced by the interplay of decay and $B_d^0 - \bar{B}_d^0$ mixing. This kind of $CP$-violating signals can be described by the following rephasing-invariant quantity:

$$\xi_f \;=\; \mathrm{Im}\left[\frac{q_B}{p_B}\,\frac{A(\bar{B}_d^0 \to f)}{A(B_d^0 \to f)}\right] \;. \tag{22}$$

For $B_d \to \psi K_S$ in the minimal standard model, the penguin contribution is expected to be negligibly small and $K^0 - \bar{K}^0$ mixing in the final state leads to an additional weak phase $(V_{cd}^* V_{cs})/(V_{cd} V_{cs}^*)$. These two arguments may also be valid for either of the above two models with an extended quark sector [12, 22]. Therefore, one can obtain

$$\xi_{\psi K_S} \;=\; \mathrm{Im}\left[-\frac{V_{tb}^* V_{td}}{V_{tb} V_{td}^*}\,e^{\mathrm{i}\Phi}\cdot\frac{V_{cs}^* V_{cb}}{V_{cs} V_{cb}^*}\cdot\frac{V_{cd}^* V_{cs}}{V_{cd} V_{cs}^*}\right] \;=\; \sin(2\beta - \Phi)\;. \tag{23}$$

For $B_d \to \pi^+\pi^-$, we have

$$\xi_{\pi^+\pi^-} = \text{Im}\left[\frac{V_{tb}^*V_{td}}{V_{tb}V_{td}^*}e^{i\Phi} \cdot \frac{V_{ud}^*V_{ub}}{V_{ud}V_{ub}^*}\right] = \sin(2\alpha + \Phi), \quad (24)$$

if we neglect the penguin amplitudes. Uncertainties arising from the penguin contribution can be eliminated by use of the isospin relations among $B_d \to \pi^+\pi^-, \pi^0\pi^0$ and $B_u^\pm \to \pi^\pm\pi^0$ [13]. Although either $V_{t'b}^*V_{t'd}$ or $U_{db}$ may contaminate the penguin amplitudes, the above phase combination $(2\alpha + \Phi)$ can still be extracted from an isospin analysis. The reason is that the $I = 2$ amplitude of $B \to \pi\pi$ only contains the tree-level quark diagrams with $V_{ud}^*V_{ub}$ or $V_{ud}V_{ub}^*$, and all other weak and strong phases can be absorbed into a set of complex parameters which are determinable from the isospin triangles.

From Eqs. (23) and (24) we find

$$\arcsin(\xi_{\psi K_S}) + \arcsin(\xi_{\pi^+\pi^-}) = 2(\alpha + \beta) = 360^0 - 2\gamma, \quad (25)$$

irrelevant to the weak phase $\Phi$. Thus $\gamma$ could be extracted from the $CP$ asymmetries of $B_d \to \psi K_S$ and $\pi^+\pi^-$, even though $B_d^0 - \bar{B}_d^0$ mixing is contaminated due to the presence of new physics.

## F.   On Testing Unitarity of the $3 \times 3$ CKM Matrix

Finally we comment briefly on tests of unitarity of the $3 \times 3$ CKM matrix. From current experimental constraints on various non-standard electroweak models, we know that no new physics can significantly affect direct decays of $b$ quark. In addition to $D^0 - \bar{D}^0$ mixing, $B^0 - \bar{B}^0$ mixing and loop-induced penguin transitions are two possible places to accommodate new physics beyond the minimal standard model. Thus the $CP$ asymmetries of $B$ decays, induced either by the interplay of decay and $B^0 - \bar{B}^0$ mixing or by penguin diagrams, could be contaminated by new physics. This leads to some difficulties for us to determine a specific weak phase cleanly and to test the unitarity conditions in Eq. (1) meaningfully.

(1) First of all, the normalization relations of unitarity in Eq. (1) can be well checked with the help of more precise data on $|V_{i\alpha}|$ ($i = u, c$ and $\alpha = d, s, b$) and on $|V_{tb}|$. Among these seven matrix elements, it is urgent to minimize the experimental (and theoretical) errors associated with the values of $|V_{ub}|$, $|V_{cd}|$, $|V_{cs}|$ and $|V_{cb}|$. A determination of $|V_{tb}|$ will be available from the top-quark lifetime. At least, one can check the following three conditions:

$$\begin{aligned}|V_{ud}|^2 + |V_{us}|^2 + |V_{ub}|^2 &= 1 \quad (?), \\ |V_{cd}|^2 + |V_{cs}|^2 + |V_{cb}|^2 &= 1 \quad (?), \\ |V_{ub}|^2 + |V_{cb}|^2 + |V_{tb}|^2 &= 1 \quad (?).\end{aligned} \quad (26)$$

A clean extraction of $|V_{td}|$ and $|V_{ts}|$ from direct production or decays of the top quark will be very difficult in experiments. Although these two elements can in principle be determined from $B_d^0 - \bar{B}_d^0$ and $B^0_s - \bar{B}^0_s$ mixings respectively, this approach itself could be affected by unknown new physics. To test the validity of Eq. (26) up to $O(\lambda^5)$ or $O(\lambda^6)$, of course, much effort is needed to make.

(2) Within the $3 \times 3$ CKM scheme, the smallness of $\mathcal{A}_1$ is governed by $\lambda^6$ and it is insensitive to the exact values of other Wolfenstein parameters. This implies that $|V_{us}| - |V_{cd}| \sim \lambda^5 < 10^{-3}$ is a reliable constraint that can now be obtained from unitarity. However, the difference between the experimental central values of $|V_{us}|$ and $|V_{cd}|$ is about 0.0165, significantly larger than the above unitarity restriction. Since the relative errors associated with $|V_{us}|$ and $|V_{cd}|$ are about 0.8% and 8.3% respectively [1], we expect that more precise measurements should enhance the existing value of $|V_{cd}|$ and lead it to approach $|V_{us}|$ closely. Within the accuracy of 0.1% for both elements, a clear deviation of $|V_{cd}|$ from $|V_{us}|$ would imply unitarity breaking in the $3 \times 3$ CKM matrix. From Eqs. (2) and (7), we notice that within the $3 \times 3$ CKM scheme $|V_{ud}|$ is larger than $|V_{cs}|$ and their difference is of the order $\lambda^4 \sim 10^{-3}$. The current experimental data give $|V_{ud}| = 0.9744 \pm 0.0010$ and $|V_{cs}| = 1.01 \pm 0.18$ [1]. Of course, the precision associated with $|V_{cs}|$ is very unsatisfactory and need be improved in the forthcoming experiments [23]. Within the accuracy of 1%, the measured value of $|V_{cs}|$ should be indistinguishable from that of $|V_{ud}|$, as required by unitarity.

(3) In comparison with $\alpha$ and $\beta$, $\gamma$ could play an interesting role in testing unitarity of the $3 \times 3$ CKM matrix. Among six unitarity triangles in Fig. 1, only triangle [t] is determinable from measurements of the six matrix elements in the first two rows of $V$. If its three sides can be constrained up to the accuracy of $O(\lambda^5)$, then the angle $\gamma$ (i.e., $\angle 7$) are calculable through

$$\gamma = \arccos\left(\frac{|V^*_{ud}V_{cd}|^2 + |V^*_{ub}V_{cb}|^2 - |V^*_{us}V_{cs}|^2}{2 \; |V^*_{ud}V_{cd}| \; |V^*_{ub}V_{cb}|}\right) . \qquad (27)$$

We have observed in the preceding sections that $\varphi = \gamma_0 \approx \gamma$ for the minimal standard model. Beyond the $3 \times 3$ CKM scheme, $\gamma$ is expected to be extracted from $CP$ asymmetries in some $B$-meson decays, as illustrated in Eqs. (23 - 25). In this case, we may generally have $\varphi \neq \gamma$. Thus a comparison between the values of the relevant weak phases (e.g., $\gamma$, $\gamma_0$, $\angle 7$ and $\varphi$) obtained from different approaches should be able to check the orthogonality conditions in Eq. (1).

I would like to thank H. Fritzsch for his warm hospitality and constant encouragement. My gratitude goes also to C. Jin, D.M. Kaplan, H. Simma and D.D. Wu for useful discussions. I owe a great debt to A. Fridman and J.P. Engel for their financial support, so that I was able to participate in this nice workshop in this nice city. My research was supported by the Alexander von Humboldt Foundation of Germany.